\documentclass[11pt, doublespace, draftclsnofoot, onecolumn, A4paper]{IEEEtran}
\usepackage{amsmath}
\usepackage{amsmath,amsthm}
\usepackage{cite}   
\usepackage{graphicx}  
\usepackage{amsmath, amsopn, psfrag, amsthm}   
\usepackage{amssymb}
\usepackage{color}

\usepackage{algorithm}

\usepackage{algorithmic}
\usepackage[utf8]{inputenc}
\usepackage[english]{babel}

\usepackage{psfrag}
\usepackage{epsfig}
\graphicspath{{./img/}}
\DeclareGraphicsExtensions{.pdf,.jpeg,.png, .eps}
\DeclareMathOperator*{\argmax}{arg\, max}

\usepackage{amssymb,amsmath,mathtools}

\usepackage{cite}   
\usepackage{graphicx}  
\usepackage{amsmath, amsopn}   
\usepackage{amssymb}
\usepackage{color}
\usepackage{psfrag}
\usepackage{epsfig}
\usepackage{stfloats}
\usepackage{lipsum}
\usepackage{multicol}
\graphicspath{{./img/}}
\DeclareGraphicsExtensions{.pdf,.jpeg,.png}
\vspace{-.8cm}

\begin{document}
\setlength{\columnsep}{.261in}
\title{Robust User Scheduling with COST 2100 Channel Model for Massive MIMO Networks}
\author{Manijeh Bashar,~\IEEEmembership{Student Member,~IEEE}, Alister G. Burr,~\IEEEmembership{Member,~IEEE},  Katsuyuki Haneda,~\IEEEmembership{Member,~IEEE}, and Kanapathippillai Cumanan,~\IEEEmembership{Member,~IEEE}
\thanks{M. Bashar, A. G. Burr, and K. Cumanan  are with the Department of Electronic Engineering, University of York, Heslington, York, UK. Email: \{mb1465, alister.burr, kanapathippillai.cumanan\}@york.ac.uk.} \thanks{K. Haneda is with Aalto
University School of Electrical Engineering, Espoo, Finland. Email: {katsuyuki.haneda@aalto.fi}}}
%
\maketitle
%
%
\begin{abstract}
This paper considers a Massive multiple-input multiple-output (MIMO) network, where the base station (BS) with a large number of antennas communicates with a smaller number of users. The signals are transmitted using frequency division duplex (FDD) mode. The problem of user scheduling with reduced overhead of channel estimation in the uplink of Massive MIMO systems has been investigated. We consider the COST 2100 channel model. In this paper, we first propose a new user selection algorithm based on knowledge of the geometry of the service area and of location of clusters, without having full channel state information (CSI) at the BS. We then show that the correlation in geometry-based stochastic channel models (GSCMs) arises from the common clusters in the area. In addition, exploiting the closed-form Cramer-Rao lower bounds (CRLB)s, the analysis for the robustness of the proposed scheme to cluster position errors is presented. It is shown by analysing the capacity upper-bound that the capacity strongly depends on the position of clusters in the GSCMs and users in the system. Simulation results show that although the BS receiver does not require the channel information of all users, by the proposed geometry-based user scheduling (GUS) algorithm the sum-rate of the system is only slightly less than the well-known greedy weight clique
(GWC) scheme \cite{SUSGoldsmithGlobcom,ITC09_Userselection_GWC}. {Finally, the robustness of the proposed algorithm to cluster localization is verified by the simulation results.}
\vspace{.12cm}

\end{abstract}
\section{Introduction}
 \let\thefootnote\relax\footnotetext{The work of A. G. Burr and K. Cumanan was supported by H2020- MSCA-RISE-2015 under grant number 690750. The work on which this paper is based was carried out in collaboration with COST Action CA15104 (IRACON).}
Massive multiple-input multiple-output (MIMO) is a promising
technique to achieve high data rate \cite{5gdebbah,massivetddMarzetta14}. However, high performance multiuser MIMO (MU-MIMO) uplink techniques rely on the availability of full channel state information (CSI) of all user terminals at the base station (BS) receiver, which presents a major challenge to their practical implementation. This paper considers an uplink multiuser system where the BS is equipped with $M$ antennas and serves $K_s$ decentralized single antenna users ($M\gg K_s$). In the uplink mode, the BS estimates the uplink channel and uses linear receivers to separate the transmitted data. The BS receiver uses the estimated channel to implement the zero-forcing (ZF) receiver which is suitable for Massive MIMO systems \cite{MarzettaMRC13}. To investigate the performance of MIMO systems, an accurate small scale fading channel model is necessary. 

Most standardized MIMO channel models such as IEEE $802.11$, the 3GPP spatial model, and the COST 273 model rely on clustering \cite{standard}. Geometry-based stochastic channel models (GSCMs) are mathematically tractable models to investigate the performance of MIMO systems \cite{Molish_tufvesson}. The concept of clusters has been introduced in GSCMs to model scatterers in the cell environments \cite{Molish_tufvesson}. In \cite{Rappaport_globe15}, the authors use clusters to characterize an accurate statistical spatial channel model (SSCM) in millimeter-wave (mmWave) bands by grouping multipath components (MPCs) into clusters. {MmWave communication suffers from very large path losses, and hence requires large antenna arrays in compensation.} \cite{Molish_tufvesson}. This paper investigates the throughput in the uplink for the Massive MIMO with carrier frequency in the order of 2 GHz, but the principles can also apply to other frequency bands, including mmWave.

Most existing Massive MIMO techniques rely on the availability of the full CSI of all users at the BS, which presents a major challenge in implementing Massive MIMO. As a result, Massive MIMO techniques with reduced CSI requirement are of great interest. An important issue in Massive MIMO systems is investigating user scheduling in which multiuser diversity gain with imperfect CSI is considered \cite{Liu_selection}. Recently, a range of user scheduling schemes have been proposed for large MIMO systems. Most of these, such as that described in \cite{Lee14user}, require accurate knowledge of the channel from all potential users to the BS -which in the frequency division duplex (FDD) Massive MIMO case is completely infeasible to obtain. In \cite{XuFDDuser}, the authors proposed a greedy user selection scheme by exploiting the instantaneous CSI of all users. However, in this paper we focus on a simplified and robust user scheduling algorithm, by considering Massive MIMO simplifications and the effect of the cell geometry.
\subsection{Contributions of This Work}
This work investigates a new user selection algorithm for high frequency stochastic geometry-based channels with large numbers of antennas at the BS receiver. We investigate user scheduling by considering the Massive MIMO assumption. The proposed geometry-based user scheduling (GUS) is similar to the greedy weight clique
(GWC) algorithm but with a different cost function. In the GUS algorithm, the BS selects users based only on the geometry of the area, while in the GWC, the BS uses the channel of the users for user scheduling. Given a map of the area of the micro-cell, we perform efficient user scheduling based only on the position of users and clusters in the cell. In GSCMs, MPCs from common clusters cause high correlation which reduces the rank of the channel. In this paper, we investigate the effect of common clusters on the system performance. Moreover, we assume that the space-alternating generalized expectation (SAGE) algorithm \cite{crlbc,crlbj} is used (offline) to estimate the direction of arrival (DoA) and the delay of the path. The performance analysis shows the significant effect of the distinct clusters on the system throughput. We prove that to maximize the capacity of system, it is required to select users with visibility of the maximum number of distinct clusters in the area. Next, we show that the position of clusters in the area can be given by geometrical calculation. 
Our results and contributions are summarized as follows: 

\begin{itemize}
\item
 Close analytical approximations for Massive MIMO systems are found.
\item
 Using the map of the area and positions of users, a new user scheduling scheme is proposed \textit{under the assumption of no CSI at the BS, other than the location of clusters}.
 Since the position of clusters in the area are fixed, we assume that cluster localization can be done offline. 
\item 
 Simulation results show that the proposed scheme significantly reduces the overhead channel estimation in Massive MIMO systems compared to conventional user scheduling algorithms, especially for indoor and outdoor of micro-cells.
\item
To investigate the robustness of the proposed algorithm to cluster localization, the performance degradation is shown for different values of the error in cluster localization and simulation results show the robustness of the proposed user scheduling algorithm to poor cluster localization.
\end{itemize}

\subsection{Outline}
The rest of the paper is organized as follows. Section II describes
the system model. The proposed user scheduling scheme is presented in Section III. Section IV presents performance analysis of the
proposed user scheduling with no estimated CSI. {The robustness of the proposed user scheduling algorithm  to cluster localization errors is investigated in Section V.} Numerical
results are presented in Section VI. Finally, Section VII concludes
the paper. 
\subsection{Notation}
Note that in this paper, uppercase and lowercase
boldface letters are used for matrices and vectors, respectively.
The notation $\mathbb{E}\{\cdot\}$ denotes expectation. Moreover, $|\cdot|$ stands for absolute value. Conjugate
transpose of vector $\textbf{x}$ is $\textbf{x}^{H}$. Finally $\textbf{x}^T$ and $\textbf{X}^{\dag}$
denote the
transpose of vector $\textbf{x}$ and the pseudo-inverse of matrix $\textbf{X}$, respectively.
\section{SYSTEM MODEL}
We consider uplink transmission in a single cell Massive MIMO system with $M$ antennas at the BS and $K> M$ single antenna users. The $M \times 1$ received signal at the BS when $K_s\, (K_s\ll M)$ users have been selected from the pool of $K$ users, is given by 
\begin{equation}
\textbf{r}= \sqrt{p_k}\textbf{H}\textbf{x}+n, 
\end{equation}
where $\textbf{x}$ represents the symbol vector of $K_s$ users, $p_k$ 
is the average power of the $k$th user and $\textbf{H}$ denotes the aggregate $M \times K_s$ channel of all selected users. The BS is assumed to have CSI only of the selected users \cite{my-master-vtc,myiet_master}. We are interested in a linear ZF receiver which can be provided by evaluating the pseudo-inverse of $\textbf{H}$, the aggregate channel of all selected users according to 
\begin{equation}
\textbf{W}=\textbf{H}^{\dag}=\left(\textbf{H}^{H}\textbf{H}\right)^{-1}\textbf{H}^{H}.
\end{equation}
 Then after using the detector, the received signal at the BS is
\begin{equation}
\textbf{y}= \sqrt{p_k}\textbf{W}\textbf{H}\textbf{x}+\textbf{W}\textbf{n}.
\end{equation}
Let us consider equal power allocation between users, i.e. $p=\frac{P_t}{k}$, in which $P_t$ denotes the total power. The achievable sum-rate of the system is obtained as \cite{my-master-vtc,myiet_master}
\begin{equation}
R=\sum_{k=1}^{K_s}{\log_{2}
 \bigg({{  1+\frac{p|{\mathbf{w}_k}{\mathbf{h}_k}|^2}{1+\sum_{i=1,i\ne{k}}^{K}p|\mathbf{w}_k\mathbf{h}_i|^2}}}\bigg)},
\end{equation}
where $\textbf{w}_k$ and $\textbf{h}_k$ are respectively the $k$th rows of the matrix $\textbf{W}=[\textbf{w}_1^T,\textbf{w}_2^T,\cdots,\textbf{w}_{K_s}^T]^T$, and  the \textit{k}th column of  $\textbf{H}=[\textbf{h}_1,\textbf{h}_2,\cdots,\textbf{h}_{K_s}]$.
\subsection{Geometry-based Stochastic Channel Model}

\begin{figure}
\center
\includegraphics[width=91mm]{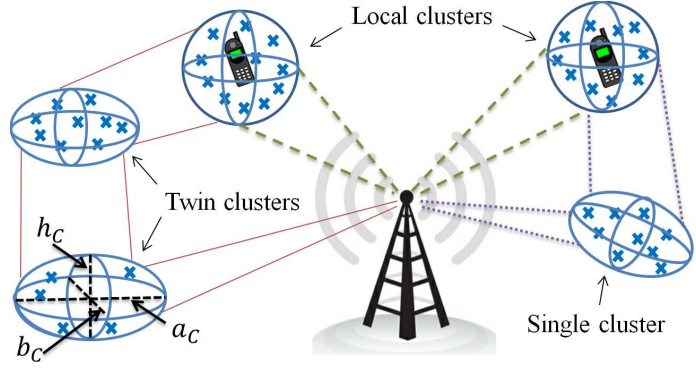}
\caption{The general description of the cluster model. The spatial spreads for $c$th cluster are given.}
\label{cluster}
\end{figure}
In GSCMs, the double directional channel impulse response is a superposition of MPCs. The channel is given by \cite{Costaction}
\begin{equation}
h(t,\tau,\phi,\theta)=\sum_{j=1}^{N_C}\sum_{i=1}^{N_p}a_{i,j}\delta(\phi-\phi_{i,j})\delta(\theta-\theta_{i,j})\delta(\tau-\tau_{i,j}),
\label{h1}
\end{equation}
where $N_p$ denotes the number of multipath components, $t$ is time, $\tau$ denotes the delay, $\delta$ denotes the Dirac delta function, and $\phi$ and $\theta$ represent the direction of arrival (DoA) and direction of departure (DoD) respectively. Similar to \cite{Costaction,our_ew}, we group the multipath components with similar delay and directions into clusters. Three kinds of clusters are defined; local clusters, single clusters and twin clusters. Local clusters are located around users and the BS while single clusters are represented by one cluster and twin clusters are characterized by two clusters related respectively to the user and BS side as shown in Fig. \ref{cluster}. A local cluster is a single cluster that surrounds a user: single clusters can also occur in a different position. Twin clusters consist of a linked pair of clusters, one of which defines the angles of departure of multipaths from the transmitter, while the other defines the angles of arrival at the receiver \cite{Costaction}. There is a large number of clusters in the area, however just some of them can contribute to the channel. The circular visibility region (VR) determines whether the cluster is active or not for a given user. The MPC's gain scales by a transition function that is given by 
\begin{equation}
A_{VR}(\bar{r}_{MS})=\dfrac{1}{2}-\dfrac{1}{\pi}\arctan \left(\dfrac{2\sqrt{2}\left(L_c+d_{MS,VR}-R_C\right)}{\sqrt{\lambda L_c}}\right),
\label{vr}
\end{equation}
where $\bar{r}_{MS}$ is the centre of the VR, $R_C$ denotes the VR radius, $L_C$ represents the size of the transition region and $d_{MS,VR}$ refers to the distance between the mobile stations (MS)s and the VR centre. For a constant expected number of clusters $N_C$, the area density of VRs is given by
\begin{equation}
\rho_C=\frac{N_C-1}{\pi \left(R_C-L_C\right)^2}.
\label{rho_C}
\end{equation}
All clusters are ellipsoids in the environment and can be characterized by the cluster spatial delay spread, elevation spread and azimuth spread. Once the position of the BS and users are fixed, we need to determine the positions of the clusters in the area by geometrical calculations. For the local clusters, we consider a circle around the users and  the BS, so that the size of the local cluster can be characterized by the cluster delay spread ($a_C$), elevation spread ($h_C$) and the position of MPCs \cite{Costaction}. For local clusters the cluster delay, azimuth and elevation spreads can be given by 
\begin{subequations}
\begin{eqnarray}
&a_C=\dfrac{\Delta\tau c_0}{2},\\
& b_C=a_C,\\
& h_C=d_{C,BS}\tan \theta_{BS},
\end{eqnarray}
\end{subequations}
where $c_0$ denotes the speed of light, $d_{C,BS}$ is the distance between the cluster and the BS, $\Delta\tau$ refers to the delay spread and $\theta_{BS}$ is the elevation spread seen by the BS. The delay spread, angular spreads and shadow fading are correlated random variables and for all kinds of clusters are given by \cite{corria}
\begin{subequations}
\begin{eqnarray}
& \Delta\tau_c=\mu_\tau(\frac{d}{1000})^{\frac{1}{2}} 10^{\sigma_\tau \frac{Z_c}{10}},\\
& \beta_c= \tau_\beta 10^{\sigma_\beta \frac{Y_c}{10}},\\
& S_m = 10^{\sigma_s \frac{X_c}{10}},
\end{eqnarray}
\end{subequations}
where $\Delta\tau_c$ refers to the delay spread, $\beta_c$ denotes angular spread, and $S_m$ is the shadow fading of cluster $c$. Moreover, $X_c$, $Y_c$ and $Z_c$ denote correlated random variables with zero mean and unit variance. Correlated random process can be computed by Cholesky factorization \cite{corria}. Cholesky factorization can be used to generate a random vector with a desired covariance matrix \cite{matrixbook}. The MPCs' positions can be drawn from the truncated Gaussian distribution given by \cite{Costaction}
\begin{equation}
f(r)\!=\!\left\{
\begin{array}{rl}\dfrac{1}{\sqrt{2\pi\sigma_{r,o}^2}}\exp\left(\!-\!(\dfrac{r\!-\!\mu_{r,o}}{\sqrt{2}\sigma_{r,o}})^2\right)& |r|\le {r}_{T},\\
0~~~~~~~~~~~~~~~~~~~~~~~~~~ &\text{otherwise},
\end{array} \right.
\label{fr1}
\end{equation}
where $r_T$ denotes the truncation value. For single clusters, the cluster delay, azimuth and elevation spreads can be given by 
\begin{subequations}
\begin{eqnarray}
&a_C=\Delta\tau c_0/2,\\
&b_C=d_{C,BS}\tan \phi_{BS},\\
&h_C=d_{C,BS}\tan \theta_{BS}.
\end{eqnarray}
\end{subequations}
To get the fixed positions of the single clusters, the radial distance of the cluster from the BS drawn from the exponential distribution \cite{Costaction}
\begin{equation}
f(r)=\left\{
\begin{array}{rl}
0~~~~~~~~~~~~~~~~~~~~~~~~~~ & r<r_{min},\\
\dfrac{1}{\sigma_r}\exp\left(-(r-\dfrac{r_{min}}{\sigma_r})\right)& \text{otherwise}.
\end{array} \right.
\label{fr}
\end{equation}
To determine the fixed position of the cluster, the angle of the cluster can be drawn from the Gaussian distribution with a standard deviation $\sigma_{\phi,C}$. For the twin clusters, for both the BS and user side clusters we have 
\begin{subequations}
\begin{eqnarray}
& a_C=\dfrac{\Delta\tau c_0}{2},\\
& b_C=d_{C,BS}\tan \phi_{BS}.
\end{eqnarray}
\end{subequations}
For the BS side cluster, the elevation spread can be given by 
\begin{equation}
h_C=d_{C,BS}\tan \theta_{BS},
 \end{equation}
  while for the MS side cluster, we have
\begin{equation}
h_C=d_{C,MS}\tan \theta_{MS}.
  \end{equation}
   Fig. \ref{cluster} gives an example of the geometry of the $C$th cluster. For twin clusters, the distance between the cluster and the BS and the distance from the VR center and the MS is given by \cite{Costaction}
\begin{equation}
d_{C,BS}\tan{\Phi_{C,BS}}=d_{C,MS}\tan{\Phi_{C,MS}}.
\end{equation}
The delay of  a cluster is represented by \cite{Costaction}
\begin{equation}
\tau_C={(d_{C,BS}+d_{C,MS}+d_C)} / {c_0}+\tau_{C,link},
\end{equation}
where the geometrical distance between twin clusters is represented by $d_C$, $d_{C,MS}$ denotes the geometrical distance between the user and the center of the visibility region, $d_{C,BS}$ refers to the distance between the BS and the cluster, and finally $\tau_{C,link}$ is the cluster link delay between the twin clusters. Hence, the cluster power attenuation is given by \cite{Costaction}
\begin{equation}
A_C=\max \left(\exp\left[-k_\tau (\tau_C-\tau_0) \right], \exp\left[-k_\tau (\tau_B-\tau_0) \right]\right),
\label{Ac}
\end{equation}
where $k_\tau$ denotes the decay parameter, and $\tau_B$ is the cut-off delay. We assume Rayleigh fading for the MPCs within each cluster. Hence, the complex amplitude of the $i$th MPC in the $j$th cluster in (\ref{h1}) is given by
 \begin{equation}
a_{i,j}=\sqrt{L_p}A_{VR}\sqrt{A_C A_{MPC}}~e^{-j2\pi f_c \tau_{i,j}},
\label{a}
\end{equation} 
where $L_p$ is the channel path loss, $A_{MPC}$ is the power of each MPC which is characterized by the Rayleigh fading distribution and $\tau_{i,j}$ is the delay of the $i$th MPC in cluster $j$ given by \cite{Costaction}
\begin{equation}
\tau_{i,j}=\frac{\left(d_{MPC_{i,j},BS}+d_{MPC_{i,j},MS}\right)}{c_0}+\tau_{i,C,link} .
\label{tau2}
\end{equation}
By assuming a fixed OFDM subcarrier, we can drop the variable $\tau_{i,j}$ from (\ref{a}). For the non-line-of-sight (NLoS) case of the micro-cell scenario, the path loss expression can be given by \cite{pathloss}
\begin{IEEEeqnarray}{rCl}
L = 26 \log_{10}d+20\log_{10}\left(\dfrac{4\pi}{\lambda}\right),
\label{d}
\end{IEEEeqnarray}where $d$ and $\lambda$ denote the distance (in meter) and the wavelength (in meter), respectively.\\
\section{Geometry-based User Scheduling}
In this section, we consider user scheduling with ZF based on the position of clusters and users in the area. In order to avoid a huge channel estimation load in the uplink of a Massive MIMO system with many users and antennas, we propose to estimate only the channels of the selected users. The reduction in the amount of channel estimation required between each transmit and receive antenna is the important result of the proposed scheme. The gain achieved by selecting users with the strongest channel is referred to as multiuser diversity and requires CSI of all users \cite{GoldsmithJurnal}. However, we propose a new user selection scheme which relies on maximizing the number of distinct clusters seen by the scheduled users. In the next subsections, we prove that the proposed scheme results in less inter-user interference (IUI) and increases the users' SINR and the system's sum-rate. In the following subsection, we present a scheme to select users which maximizes the long term sum-rate and as it is based on the position of the users and does not need the estimated channel of all users in the uplink, and hence can be a practical user selection scheme for large MIMO systems. For this case, the performance analysis are found in the next subsection.
\subsection{Proposed Geometry-based User Scheduling (GUS)}
In this section, an algorithm is proposed for increasing the system throughput based on the geometry of the system and without estimating the channels of all users in the area. Once the set of active users has been determined, the receiver BS estimates the channels of the selected users and the users transmit data. Next, the performance of the proposed user selection algorithm to maximize the sum-rate is evaluated. In large MIMO systems with large numbers of users estimating the channels of all users is practically difficult. So the proposed user scheduling algorithm can be an efficient way to reduce the overhead of channel estimation.\\   
First, we generate the matrix $\textbf{V}$, as the following 
\begin{equation}
\textbf{{V}}=
\begin{bmatrix} 
v_{1}^1 & v_{2}^1 & \ldots & {v}_{N_C}^1 \\
v_{1}^2 & {v}_{2}^2 & \ldots & {v}_{N_C}^2 \\
\vdots & \vdots & \ddots & \vdots \\
v_{1}^K & v_{2}^K & \ldots & v_{N_C}^K
\end{bmatrix},
\end{equation}
where 
\begin{equation}
v_{i}^j=\sqrt{L_{p,i}^j}A_{VR,i}^j\sqrt{A_{C,i}^j},
\label{a1}
\end{equation} 
where $L_{p}^j$ denotes the channel path loss for user $j$, $A_{VR,i}^j$ is the MPC power attenuation which is a function of the distance between the user $i$ and the centre of the visibility region related to the $j$th cluster and is given by (\ref{vr}), and $A_{C,i}^j$ denotes the cluster power attenuation given by (\ref{Ac}) for the user $j$ and the $i$th cluster. So, the matrix $\textbf{V}$ is a function of the distance from the BS to users, the distance of the BS from clusters and from users to the center of the visibility region.

\begin{algorithm}[t!]
\caption{{\small Geometry-based User Scheduling (GUS) Algorithm}}
Step 1) Initialization:~~$\mathcal{W}_{0}=[1,\cdots,K]$, $\mathcal{S}_{0}=\emptyset$, $i=1$,\\
Step 2) Load position of users, for example by means of GPS,\\
Step 3) Generate matrix $\textbf{V}$,\\
Step 4) Greedy Algorithm:
\begin{itemize}
\item{4.1}\begin{small}
$\pi(i)=\operatornamewithlimits{argmax}\limits_{k \in \mathcal{W}_{0}}f_1(||\textbf{v}_k||)=\operatornamewithlimits{argmax}\limits_{k \in \mathcal{W}_{0}}||\textbf{v}_k||$,\\ $\textcolor{blue}{\mathcal{S}_{0}\gets \mathcal{S}_{0}\cup \{\pi(i)\}}$, $\hat{\textbf{v}}_{(i)}=\textbf{v}_{(\pi(i))}$,\\
\end{small}
\item{4.2}~If $|\mathcal{S}_{0}|<K_s$, $\mathcal{W}_{i}=\{k \in \mathcal{W}_{i-1}, k \ne \pi(i) \}$,\\
\item{4.3}
\begin{small}
$\pi(i)=\operatornamewithlimits{argmin}\limits_{k \in \mathcal{W}_{i-1}}f_2(\textbf{v}_k,\hat{\textbf{v}}_{(i)})=\operatornamewithlimits{argmin}\limits_{k \in \mathcal{W}_{i-1}}\frac{  |\textbf{v}_k\hat{\textbf{v}}_{(i)}^{*}|  }{  ||\textbf{v}_k||||\hat{\textbf{v}}_{(i)}||  } \}$,~$\mathcal{S}_{0}\gets \mathcal{S}_{0}\cup \{k\}$, $\hat{\textbf{v}}_{(i)}=\textbf{v}_{(\pi(i))}$,
\end{small}
\item{4.4}~$i=i+1$
\end{itemize}
Step 5) The BS estimates the channels of the selected users.
\end{algorithm}

\begin{algorithm}[t!]
\caption{\small Geometry-based User Scheduling (GUS) Algorithm}
\label{aluserselectiond}
\hrulefill

\textbf{1.} Initialize
$\mathcal{W}_0=[1,\cdots,K]$, $\mathcal{S}_0=\emptyset$, $i=1$

\textbf{2.} Repeat until $|\mathcal{S}_{0}|=K_s$

\textbf{3.} $i=i+1$

\textbf{4.} $\pi(i)=\argmax_{k \in \mathcal{W}_{i-1}}f(||\textbf{v}_{k}||)=\argmax_{k \in \mathcal{W}_{i-1}}||\textbf{v}_{k}||$, \\
$\mathcal{S}_{0}\gets \mathcal{S}_{0}\cup \{\pi(i)\}$, $\hat{\textbf{v}}_{(i)}=\textbf{v}_{(\pi(i))}$
\\
\textbf{5.} $\mathcal{W}_{i}=\{k \in \mathcal{W}_{i-1}, k \ne \pi(i), \frac{|\textbf{v}_{k}\hat{\textbf{v}}_{(i)}^{T}|}{||\textbf{v}_{k}||||\hat{\textbf{v}}_{(i)}||}<\epsilon_h \}$

\textbf{6.} If $|\mathcal{W}|=0$, end
\end{algorithm}

Note that increasing $\epsilon_h$ allows the users to have a larger number of shared clusters. If the value of $\epsilon_h$ is too high, Algorithm selects users with a large normalized correlation which can reduce the sum rate due to the interference in the number of selected users. For a low $\epsilon_h$, the number of users in set $\mathcal{V}_0$ in step 5
decreases and Algorithm \ref{aluserselectiond} selects a small number of users.
{Suppose $W_{0}$ contains user indices considered in the proposed algorithm.  Finally, $S_{0}$ contains $K_s=|S_{0}|$ indices of the selected users.}  
\begin{figure}[t!]
\center
\includegraphics[width=75mm]{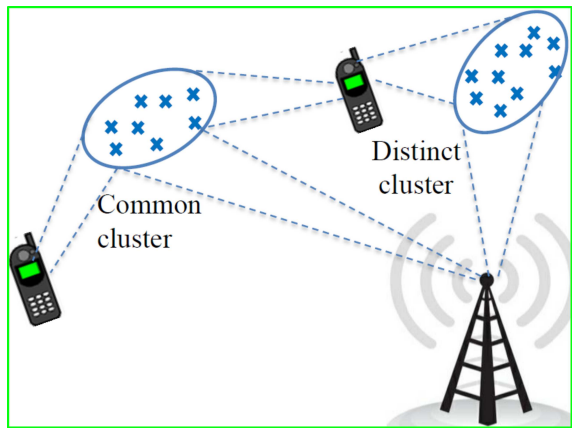}
\caption{As example of users common cluster which causes correlation.}
\label{common}
\end{figure}
\section{Performance Analysis}
If perfect CSI is available at the BS, and assuming Gaussian input, the ergodic capacity is given by
\begin{IEEEeqnarray}{rCl}
C = \mathbb{E} \left\{\log_2\det  \left(\textbf{I}+\dfrac{P_t}{K_s}\textbf{H}\textbf{H}^H\right)\right\},
\end{IEEEeqnarray}
where the term $\frac{P_t}{K_s}$ is due to the equal-power allocation, $\textbf{I}$ is an identity matrix and the channel matrix is given at the bottom of this page, where $\textit{C}(K)$ denotes the clusters seen by the $k$th user and $\alpha=-2\pi\frac{d}{\lambda}$, where $d$ denotes the spacing between two antenna elements. In GSCMs, common clusters can reduce the rank of the channel and the capacity of the system, especially at finite SNR. These common clusters also affect the multiplexing gain of the system. {Fig. \ref{common} illustrates the concept of common and distinct clusters.} When the number of objects is less than the number of BS antennas and all objects are shared between the users, achieving maximum multiplexing gain is impossible \cite{Alister10ISWCS,burrijas}. 

For ease of mathematical tractability, we analyse the capacity of a correlated three-user uplink using an upper bound. In the case of a large number of antennas at the BS, the capacity upper bound can be achieved in the case of distinct clusters. Note that in the case of a large number of transmit antennas, the elements of $\textbf{H}\textbf{H}^H$ converge to the correlation matrix so that $\textbf{R}\approx\textbf{H}\textbf{H}^H$. Hence, we have
\begin{IEEEeqnarray}{rCl}
\label{Capp}
C&=&\mathbb{E}\left\{\log_2\det\left(\textbf{I}+\dfrac{P_t}{K}\textbf{H}^H\textbf{H}\right)\right\}\nonumber \\
&\approx&\log_2\det \left(\textbf{I}+\dfrac{P_t}{K}\textbf{R}\right),
\end{IEEEeqnarray}
\begin{figure*}[!b]
\begin{small}
\hrulefill
\begin{align}
\begin{split}
\!\!\textbf{H}\!\!=\!\!\!\!
\begin{bmatrix} 
\sum_{j\in \textit{C}(1)}\sum_{i=1}^{N_p}a_{i,j} & \sum_{l\in \textit{C}(2)}\sum_{i=1}^{N_p}a_{i,l}  & \ldots & \!\!\sum_{m\in \textit{C}(K)}\sum_{i=1}^{N_p}a_{i,m} \\
\sum_{j\in \textit{C}(1)}\sum_{i=1}^{N_p}a_{i,j}e^{j\alpha\sin\phi_{i,j}} & \sum_{l\in \textit{C}(2)}\sum_{i=1}^{N_p}a_{i,l}e^{j\alpha\sin\phi_{i,l}} & \ldots & \!\!\sum_{m\in \textit{C}(K)}\sum_{i=1}^{N_p}a_{i,m}e^{j\alpha\sin\phi_{i,m}} \\
\vdots & \vdots & \ddots & \vdots \\
\sum_{j\in \textit{C}(1)}\sum_{i=1}^{N_p}a_{i,j}e^{j\alpha(M-1)\sin\phi_{i,j}} & \sum_{l\in \textit{C}(2)}\sum_{i=1}^{N_p}a_{i,l}e^{j\alpha(M-1)\sin\phi_{i,l}} & \ldots & \!\!\sum_{m\in \textit{C}(K)}\sum_{i=1}^{N_p}a_{i,m}e^{j\alpha(M-1)\sin\phi_{i,m}}\!\!\!
\end{bmatrix}\!\!,
\label{H3}
\end{split}
\end{align}
\end{small}
\end{figure*}
where $R$ is the channel correlation matrix and is given by
\begin{equation}
\textbf{R}=\mathbb{E}\left\{\textbf{H}^H\textbf{H}\right\}=
\begin{bmatrix} 
1 & r_{12}  & r_{13} \\ 
r_{12}^* & 1 & r_{23} \\
r_{13}^* & r_{23}^* & 1
\end{bmatrix},
\label{R}
\end{equation}
where
\begin{subequations}
\begin{eqnarray}
&r_{12}=\mathbb{E}\left\{\textbf{h}_1^H\textbf{h}_2\right\}=\zeta_{12}e^{j\beta_{12}},\\
&r_{13}=\mathbb{E}\left\{\textbf{h}_1^H\textbf{h}_3\right\}=\zeta_{13}e^{j\beta_{13}},\\
&r_{23}=\mathbb{E}\left\{\textbf{h}_2^H\textbf{h}_3\right\}=\zeta_{23}e^{j\beta_{23}}.
\end{eqnarray}
\end{subequations}
The term $r_{12}$ can be given by
\begin{small}
\begin{IEEEeqnarray}{rCl}
r_{12}\!&\!=\!&\!\mathbb{E}\left\{\textbf{h}_1^H\textbf{h}_2\right\}= \mathbb{E}\Big\{\sum_{j\in \textit{C}(1)}\sum_{i=1}^{N_p}a_{i,j} \sum_{l\in \textit{C}(2)}\sum_{g=1}^{N_p}a_{g,l}^* \\&
+&\sum_{j\in \textit{C}(1)}\sum_{i=1}^{N_p}a_{i,j}e^{j\alpha\sin\phi_{i,j}}\sum_{l=1\in\textit{C}(2)}\sum_{g=1}^{N_p}a_{g,l}^*e^{-j\alpha\sin\phi_{g,l}}+\ldots\!+\! \nonumber\\&  
&\sum_{j\in \textit{C}(1)}\sum_{i=1}^{N_p}a_{i,j}e^{j\alpha(M-1)\sin\phi_{i,j}}\sum_{l\in\textit{C}(2)}\!\sum_{g=1}^{N_p}a_{g,l}^*e^{-j\alpha(M-1)\sin\phi_{g,l}}\Big\}\nonumber,
\label{r1}
\end{IEEEeqnarray}
\end{small}
where $a_{i,j}$, the amplitude of the $i$ MPCs in cluster $j$, is given by (\ref{a}), and the terms $r_{13}$ and $r_{23}$ can be derived in the same way. By substituting the terms $r_{12}$, $r_{13}$ and $r_{23}$ into (\ref{Capp}), the capacity maximization problem  in a three-user scenario can be formulated as 
\begin{IEEEeqnarray}{rCl}
\label{maxc}
C\!&=\!&\!\!\!\max_{\zeta_{12},\zeta_{13},\zeta_{23}\beta_{12},\beta_{13},\beta_{23}}\!\log_2\!\Big[(1+p)^3-p^2\left(\zeta_{12}^2\!+\!\zeta_{13}^2\!+\!\zeta_{23}^2\right)\\&
&+\!p^3\Big(2\zeta_{12}\zeta_{13}\zeta_{23}\sin(\beta_{12}\!-\!\beta_{13}+\beta_{23})\!-\!\zeta_{12}^2-\zeta_{13}^2-\zeta_{23}^2\Big)\Big]\nonumber,
\end{IEEEeqnarray}
where $p=\frac{P_t}{K}$. To maximize (\ref{maxc}), the gradient search (GS) method results in $\zeta_{12}=\zeta_{13}=\zeta_{23}=0$, for different values of $\beta_{12}$, $\beta_{13}$ and $\beta_{23}$, which is the case when common clusters do not occur between the users in the cell. In the case of distinct clusters between user $m$ and user $n$, we have 
\begin{IEEEeqnarray}{rCl}
\zeta_{mn} = \mathbb{E}\left\{\sum_{j\in \textit{C}(n)}\sum_{i=1}^{N_p}a_{i,j} \sum_{l\in \textit{C}(m)}\sum_{g=1}^{N_p}a_{g,l}^*\right\}=0.
 \label{E}
\end{IEEEeqnarray}
The equation (\ref{E}) yields $\zeta_{mn}=0$, which maximizes the capacity given by (\ref{maxc}).

For the case of Massive MIMO systems with a large number of users in the cell, having distinct clusters for all users is practically difficult. In a real scenario, it is not possible to force the term $\zeta_{mn}$ zero. The proposed user scheduling algorithm selects users which do not have common clusters and consequently forces the variable $\zeta_{mn}$ to be small. A threshold is set for the power of a cluster to be considered active. In the COST channel model, each user interacts with several clusters in the area and the cluster power depends on the distance between the user and the center of visibility region and also the distance between the cluster and the BS. We define a threshold which can determine the minimum power that a cluster may have relative to the total powers. As in \cite{Tufvesson_Cost_th}, we set the cluster power threshold to $0.01\times$ total power for a cluster to be active.
A cluster is shared between two users if contributes to both users, which means the cluster powers seen by the users are more than the threshold. Hence, the optimum value of $\zeta_{mn}$ can be achieved only when there is no common cluster in the cell.
Moreover, investigating the effectiveness of the proposed user scheduling scheme in the distributed Massive MIMO systems \cite{ourjournal1,ourjournal2,ouricc1,ouricc2,ourcommletter,ourvtc18} is an interesting topic for future work.
\section{Robustness of the Proposed User Scheduling Algorithm}
\subsection{Cluster Localization}
 The BS can estimate the direction of arrival \cite{handbook}, and hence the direction of the scattering objects should be available at the BS.
There is a well-known algorithm to estimate the delay, direction of arrival and the direction of departure of the channel paths; SAGE-based algorithm \cite{crlbc,crlbj}. As a result, the BS can identify the direction of the clusters which can be seen by the users in the cell area, and hence build up a map of the location of the scattering objects. The convenient tool that has overcame the challenge of making the position of the scatterers available is the use of environment maps \cite{Costaction}, which also shows how measured angles of arrival can be identified with physical objects in the environment, and hence can be located on the map. Successive interference cancellation has also been introduced in \cite{molish-tuf-position} for scattering object identification: it uses the channel impulse response peaks in the delay domain to map scatterers to 2-D coordinates.
\subsection{{Robustness}}
In order to study the robustness of the proposed algorithm to cluster localization error, we use the well-known SAGE algorithm \cite{crlbc,crlbj}, operating offline, as mentioned above. In cluster localization, we consider a receiver BS with a antenna array consisting of $M$ sensors located at a reference point \cite{crlbc,crlbj}. Moreover, we consider planar wavefronts. 
The closed-form Cramer-Rao lower bound (CRLB) for the delay, azimuth ($\nu$) and elevation ($\theta$) of the path are given by \cite{crlbc}
\begin{subequations}
\begin{eqnarray}
& CRLB(\tau) =\dfrac{1}{\gamma_O}\dfrac{1}{8\pi^2BW} \label{crlb1}\\
& CRLB(\theta) = \dfrac{1}{\gamma_O}\dfrac{M}{2\Delta \cos (\nu)} \label{crlb2}\\
& CRLB(\nu) = \dfrac{1}{\gamma_O}\dfrac{M}{2\Delta}\label{crlb3},
\end{eqnarray}
\label{crlb}
\end{subequations}
where $BW$ is the bandwidth, and
\begin{IEEEeqnarray}{rCl}
\Delta&=&4\pi^2(\dfrac{d}{\lambda})^2(\dfrac{7}{3}M_x^3-8M_x^2+\dfrac{29}{3}M_x-4),
\end{IEEEeqnarray} 
and
\begin{IEEEeqnarray}{rCl}
\gamma_O=M \times I\times N\times |f(\nu)|^2\gamma_I,
\end{IEEEeqnarray}
 where $I$ is the number of periods of the received signal, $N$ denotes the length of the used pseudonoise
(PN) sounding sequence available at the receiver and $\gamma_I$ is the SNR at the input of each antenna \cite{crlbc,crlbj}. Moreover, the antenna electric field pattern can be given by \cite{crlbc} \begin{equation}
f(\nu)=0.67+2.67\nu-6.79\nu^2+5.7\nu^3-1.71\nu^3.
\end{equation}
 The distance between the BS and cluster ($d_{BS,C}$) is given by geometrical calculation:
\begin{IEEEeqnarray}{rCl}
\left(c\tau-d_{BS,C}\right)^2 & = &\left(h_{BS}-h_{MS}+d_{BS,C}\sin(\nu)\right)^2+ \\ &  
&\left(d_{BS,MS}-d_{BS,C}\cos(\nu)\cos(\theta)\right)^2,\nonumber
\label{dis}
\end{IEEEeqnarray}
where $c$ denotes the velocity of light, $d_{BS,MS}$ is the distance between the user and the BS in $x-y$ plane, and $h_{BS}$ and $h_{MS}$ are the BS and user heights. The distance between the user and cluster is easily given by 
\begin{equation}
d_{MS,C}+d_{BS,C}=c\tau.
\end{equation}
 After the offline localization, the BS can build up the matrix $\tilde{\textbf{V}}$ at the beginning of each time-slot, as the following 
\begin{equation}
{\tilde{\textbf{V}}}=
\begin{bmatrix} 
\tilde{v}_{1}^1 & \tilde{v}_{2}^1 & \ldots & \tilde{v}_{N_C}^1 \\
\tilde{v}_{1}^2 & \tilde{v}_{2}^2 & \ldots & \tilde{v}_{N_C}^2 \\
\vdots & \vdots & \ddots & \vdots \\
\tilde{v}_{1}^K & \tilde{v}_{2}^K & \ldots & \tilde{v}_{N_C}^K
\end{bmatrix},
\end{equation}
where
\begin{equation}
\tilde{v}_{i}^j=\sqrt{L_{p}^j}\tilde{A}_{VR,i}^j\sqrt{\tilde{A}_{C,i}^j},
\label{a2}
\end{equation} 
where $\tilde{A}_{VR,i}^j$ and $\tilde{A}_{C,i}^j$ can be calculated by the distances obtained in (\ref{dis}). Finally, for the matrix \textbf{V}, the following equation holds
\begin{equation}
\textbf{V} = \tilde{\textbf{V}}+\textbf{E},
\end{equation}
where $\textbf{E}$ is due to the estimation error in cluster localization. Then, we use $\tilde{\textbf{V}}$ instead of $\textbf{V}$ in the proposed algorithm. The numerical results verifies the robustness of the proposed algorithm to this error. 
\section{Numerical Results and Discussion}
In this section, simulation results have been provided to validate the performance of the proposed schemes with different parameters.
\subsection{Simulation Parameters for COST 2100 Channel Model}
We evaluate the throughput of the system, averaging over 50 iterations. 
A square cell with a side length of $2\times R$ has been considered so that we call $R$ the cell size and also assume users are uniformly distributed in the cell. As in \cite{MarzettaMRC13}, we assume that there is no user closer than $R_{th}=0.1\times R$ to the BS. We simulate a micro-cell environment for the NLoS case and set the operating frequency $f_C=2$GHz. The external parameters and stochastic parameters are extracted from chapter 6 of \cite{corria} and chapter 3 of \cite{Costaction}. The BS and user heights are assumed to be $h_{BS}=5$ and $h_{MS}=1.5$, respectively. In (\ref{rho_C}) $N_C=3$, $R_C=50$, and $L_C=20$. Moreover, we consider $N_{P}=6$ paths per cluster.
\begin{figure}[t!]
\center
\includegraphics[width=95mm]{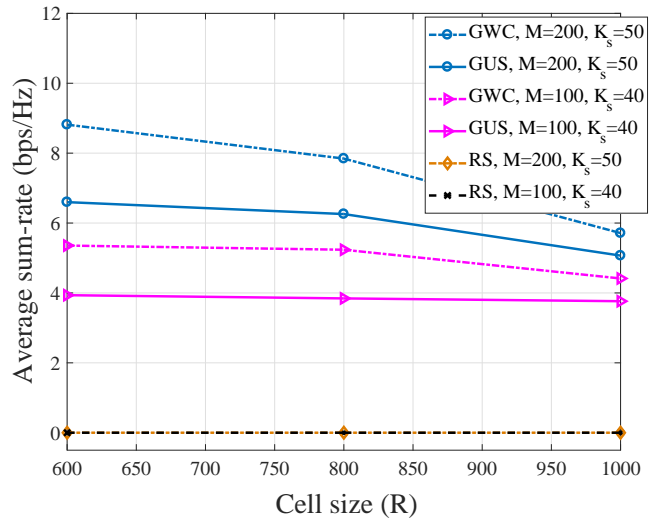}
\caption{The average sum-rate vs. the cell size for different values of $M=200$, $M=100$, $K_s=50$, and $K_s=40$. We set the total number of users in the cell $K=400$.}
\label{vsR}
\end{figure} 
\begin{figure}[t!]
\center
\includegraphics[width=95mm]{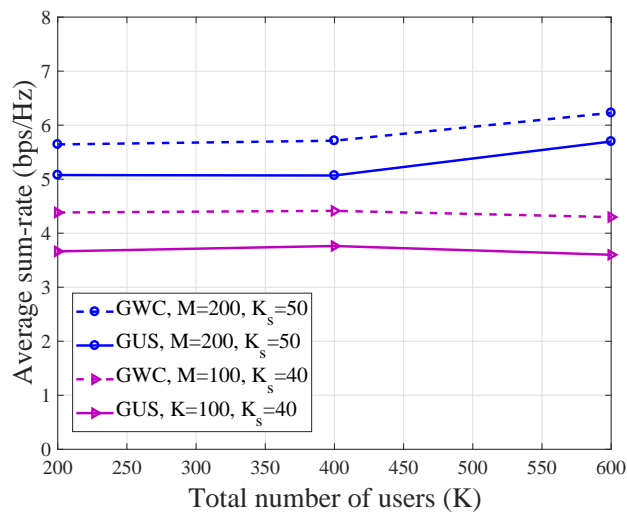}
\caption{The average sum-rate vs. total number of users for different values of $M=200$, $M=100$, $K_s=50$, $K_s=40$, $R=600$m and $R=1000$m.}
\label{vsK}
\end{figure}
\begin{figure}[t!]
\center
\includegraphics[width=95mm]{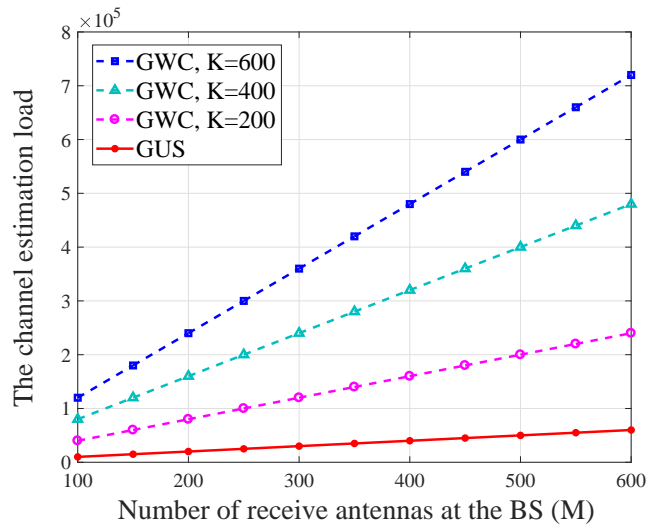}
\caption{The channel estimation load vs. value of error of antennas at the receiver BS for different values of total number of users in the cell.}
\label{vsM}
\end{figure}
\begin{figure}[t!]
\center
\includegraphics[width=95mm]{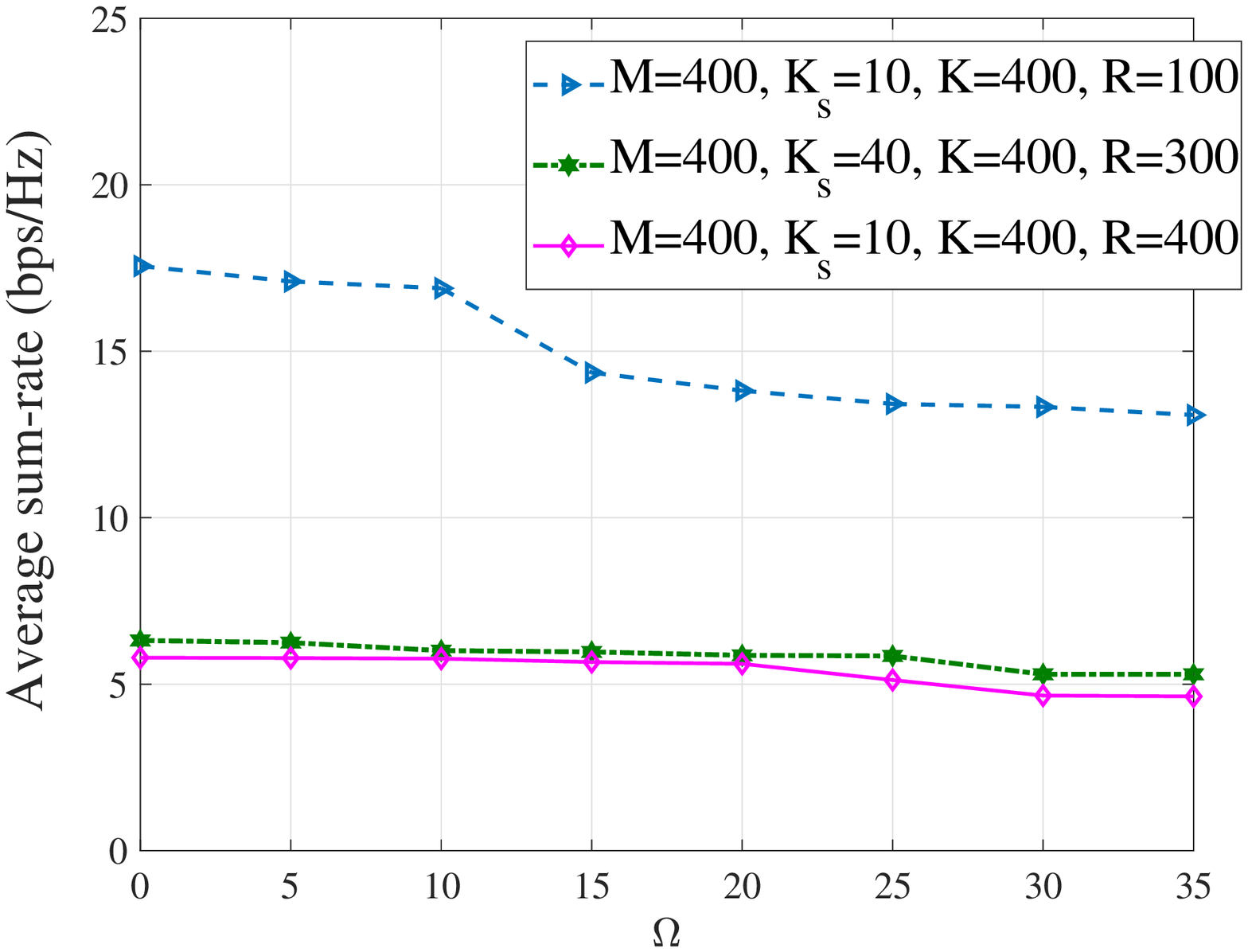}
\caption{{The average sum-rate vs. the estimation error for different values of total number of selected users in the cell and the cell size.}}
\label{e}
\end{figure}
\subsection{Simulation Results}
For this network setup, the average sum-rate is evaluated for the three scenarios. In the GUS scheme, it has been proposed that the receiver BS selects users that maximize the number of distinct clusters in the cell. We evaluate the average throughput of the GWC scheme \cite{SUSGoldsmithGlobcom,ITC09_Userselection_GWC} and random selection (RS) of users. For the case of GWC, similar to \cite{ITC09_Userselection_GWC}, we set the optimal channel direction
constraint to achieve the best performance for GWC, so the complexity of GWC is much higher than GUS.

Fig. \ref{vsR} depicts the average sum-rate with total number of receive antennas at the BS $M=100$ and $M=200$, and two values of the number of selected users $K_s=40$ and $K_s=50$ while adopting the proposed scheme with ZF receiver. As expected, since GWC exploits perfect CSI, it has the best throughput. As seen in Fig.\ref{vsR}, the performance of the proposed algorithm is slightly lower than the case in which the BS exploits full CSI and performs GWC. Interestingly, for bigger cells, the superiority of the proposed scheme is more obvious in terms of achieving performance close to that of the GWC scheme. 

In Fig. \ref{vsK}, we have plotted the average sum-rate for the case of GWC and GUS versus total number of users in the cell ($K$) with different numbers of receive antennas at the BS, $M$ and of selected users, $K_s$. In terms of average sum-rate, Fig. \ref{vsK} shows that the proposed scheme results in only a small sum-rate reduction even with a smaller total number of users.

The amount of channel estimation load required in both GWC and the proposed GUS is presented in Fig. \ref{vsM}. As the figure shows the channel estimation load of the proposed GUS is far less than that of the GWC scheme.

To investigate the robustness of the proposed scheme to different values of the error, we set 
\begin{IEEEeqnarray}{rCl}
|e|=\Omega\times \sqrt{CRLB(\rho)}, 
\end{IEEEeqnarray}
where $|e|$ denotes the absolute value of the estimation error, $\Omega$ is an integer number and $CRLB(\rho)$ is given by (\ref{crlb1})-(\ref{crlb3}), where the parameter $\rho$ can be the delay, azimuth and elevation. Fig. \ref{e} shows the average sum-rate with total number of receive antennas at the BS $M=400$, and two values of the number of selected users $K_s=10$ and $K_s=40$ versus the value of the estimation error. We set the SNR at the input of each antenna $\gamma_I=20$dB and the bandwidth $BW=20$MHz. Moreover, in equations (\ref{crlb1}) to (\ref{crlb3}), $M_x=5$, $N=127$, which are extracted from \cite{crlbc}. The figure shows the robustness of the proposed algorithm to poor cluster localization.
\section{Conclusions}
{We have investigated the user scheduling problem in Massive MIMO systems and proposed a new GUS scheme which maximizes the uplink throughput of the users, considering the FDD mode.} By applying knowledge of the location of clusters and users and the geometry of the system, we suppose that the BS does not need to estimate the channels of all users and selects users based only on the location of users and clusters in the area. {Next, exploiting Cramer-Rao lower bound, we have developed a robustness analysis for the proposed scheme.} The results show that while sum-rate slightly decreases along with the reduced overhead of channel estimation, the proposed algorithm can be an efficient scheme to reduce the complexity of user scheduling in Massive MIMO systems. {In addition, the simulation results demonstrate good robustness against the estimation error.}

\bibliographystyle{IEEEtran}
\bibliography{finalarxive} 
\end{document}